\documentclass[12pt]{article}
\usepackage[dvips]{graphicx}
\usepackage{xspace}
\usepackage{epsfig}
\usepackage{amssymb}
\usepackage{cite}
\newcommand{\Z  }[4]{\ensuremath{#1\,\pm #2\,^{+\,#3}_{-\,#4}}\xspace}

\newcommand{\aem    }{\ensuremath{\alpha}\xspace}
\newcommand{\aemsq  }{\ensuremath{\aem^2}\xspace}
\newcommand{\invpb  }{\ensuremath{\rm pb^{-1}}\xspace}
\newcommand{\qsq    }{\ensuremath{Q^{2}}\xspace}
\newcommand{\psq    }{\ensuremath{P^{2}}\xspace}

\newcommand{\lamt   }{\ensuremath{\Lambda_{\rm 3}^{\scriptstyle\overline{\rm MS}}}\xspace}
\newcommand{\ft     }{\ensuremath{F_{2}^{\gamma}}\xspace}
\newcommand{\ftn    }{\ensuremath{F_{2}^{\gamma}/\aem}\xspace}
\newcommand{\fl     }{\ensuremath{F_{\rm L}^{\gamma}}\xspace}
\newcommand{\ftxq   }{\ensuremath{\ft(x,\qsq)}\xspace}
\newcommand{\ftqn   }{\ensuremath{\ft(\qsq)/\aem}\xspace}
\newcommand{\flxq   }{\ensuremath{\fl(x,\qsq)}\xspace}

\newcommand{\ftp    }{\ensuremath{F_{2}^{\rm p}}\xspace}
\newcommand{\gsg    }{\ensuremath{\gamma^{\star}\gamma}\xspace}
\newcommand{\gev    }{\ensuremath{\rm GeV}\xspace}
\newcommand{\gevsq  }{\ensuremath{\rm GeV^2}\xspace}
\newcommand{\zn     }{\ensuremath{\rm Z^0}\xspace}
\newcommand{\epm    }{\ensuremath{\rm e^\pm}\xspace}
\newcommand{\epem   }{\ensuremath{\rm e^+e^-}\xspace}
\newcommand{\gghad  }{\ensuremath{\gsg\rightarrow{\it hadrons}}\xspace}
\newcommand{\znhad  }{\ensuremath{\zn\rightarrow{\it hadrons}}\xspace}
\newcommand{\zntau  }{\ensuremath{\zn\rightarrow \tau^+\tau^-}\xspace}
\newcommand{\gglep  }{\ensuremath{\gamma^{\star}\gamma\rightarrow{\it leptons}}\xspace}
\newcommand{\ggtau  }{\ensuremath{\gamma^{\star}\gamma\rightarrow\tau^+\tau^-}\xspace}
\newcommand{\ggel  }{\ensuremath{\gamma^{\star}\gamma\rightarrow {\rm e^+ e^-}}\xspace}
\newcommand{\eeqq   }{\ensuremath{{\rm e^+e^-}\,{\rm q}\overline{{\rm q}}}\xspace}
\newcommand{\eell   }{\ensuremath{{\rm e^+e^-}\,\ell^+\ell^-}\xspace}
\newcommand{\thet   }{\ensuremath{\theta}\xspace}
\newcommand{\ph     }{\ensuremath{\phi}\xspace}
\newcommand{\etag   }{\ensuremath{E_{\rm tag}}\xspace}
\newcommand{\ea     }{\ensuremath{E_{\rm at}}\xspace}

\newcommand{\nch    }{\ensuremath{N_{\rm ch}}\xspace}
\newcommand{\eb     }{\ensuremath{E_{\rm b}}\xspace}
\newcommand{\ttag   }{\ensuremath{\theta_{\rm tag}}\xspace}
\newcommand{\qzm    }{\ensuremath{\langle \qsq \rangle}\xspace}

\newcommand{\mc     }{\ensuremath{m_{\rm c}}\xspace}
\newcommand{\ssee   }{\ensuremath{\sqrt{s_{\rm ee}}}\xspace}
\newcommand{\pt     }{\ensuremath{p_{\rm t}}\xspace}
\newcommand{\Wvis   }{\ensuremath{W_{\rm vis}}\xspace}
\newcommand{\xvis   }{\ensuremath{x_{\rm vis}}\xspace}
\newcommand{\slop   }{\ensuremath{\aem^{-1}{\rm d}\ft/{\rm d}\ln\qsq}\xspace}
\newcommand{\slopval}{\ensuremath{0.10^{+0.05}_{-0.03}}\xspace}
\newcommand{\chidof }{\ensuremath{\chi^2/\rm dof}\xspace}
\newcommand{\act    }{\ensuremath{|\cos\theta|}\xspace}
\newcommand{\restr  }{\ensuremath{{\sigma_{\pt}}/{\pt}=\sqrt{(0.02)^2+(0.0015\,\pt)^2}}\xspace}
\newcommand{\flow  }{\ensuremath{1/N~{\rm d}E/{\rm d}\eta}\xspace}
\begin{document}
\begin{titlepage}
\begin{center}{\large   EUROPEAN LABORATORY FOR PARTICLE PHYSICS
}\end{center}\bigskip
\begin{flushright}
   CERN-PPE/97-087   \\ 18 July 1997
\end{flushright}
\bigskip\bigskip\bigskip\bigskip\bigskip
\begin{center}{\huge\bf\boldmath
 Measurement of the \qsq evolution \\ of the photon structure function \ft
}\end{center}\bigskip\bigskip
\begin{center}{\LARGE The OPAL Collaboration
}\end{center}\bigskip\bigskip
\bigskip\begin{center}{\large  Abstract}\end{center}
 New measurements are presented of the photon structure
 function \ftxq at four values of \qsq between 9 and 59~\gevsq
 based on data collected with the OPAL detector at centre-of-mass energies 
 of 161$-$172~\gev, with a total integrated luminosity of 18.1~\invpb.
 The evolution of \ft with \qsq in bins of $x$ is determined
 in the \qsq range from 1.86 to 135~\gevsq using 
 data taken at centre-of-mass energies of 91~\gev and 161$-$172~\gev.
 \ft is observed to increase with \qsq with a slope of $\slop=\slopval$
  measured in the range $0.1<x<0.6$.
\bigskip\bigskip\bigskip\bigskip
\bigskip\bigskip
\begin{center}{\large
(Submitted to Physics Letters B)
}\end{center}
\end{titlepage}
\begin{center}{\Large        The OPAL Collaboration
}\end{center}\bigskip
\begin{center}{
K.\thinspace Ackerstaff$^{  8}$,
G.\thinspace Alexander$^{ 23}$,
J.\thinspace Allison$^{ 16}$,
N.\thinspace Altekamp$^{  5}$,
K.J.\thinspace Anderson$^{  9}$,
S.\thinspace Anderson$^{ 12}$,
S.\thinspace Arcelli$^{  2}$,
S.\thinspace Asai$^{ 24}$,
D.\thinspace Axen$^{ 29}$,
G.\thinspace Azuelos$^{ 18,  a}$
% ...
}\end{center}\bigskip
\bigskip
% ...
$^{  2}$Dipartimento di Fisica dell' Universit\`a di Bologna and INFN,
I-40126 Bologna, Italy
\newline
% ...
$^{  5}$Cavendish Laboratory, Cambridge CB3 0HE, UK
\newline
% ...
$^{  8}$CERN, European Organisation for Particle Physics,
CH-1211 Geneva 23, Switzerland
\newline
$^{  9}$Enrico Fermi Institute and Department of Physics,
University of Chicago, Chicago IL 60637, USA
\newline
% ...
$^{ 12}$Indiana University, Department of Physics,
Swain Hall West 117, Bloomington IN 47405, USA
\newline
% ...
$^{ 16}$Department of Physics, Schuster Laboratory, The University,
Manchester M13 9PL, UK
\newline
% ...
$^{ 18}$Laboratoire de Physique Nucl\'eaire, Universit\'e de Montr\'eal,
Montr\'eal, Quebec H3C 3J7, Canada
\newline
% ...
$^{ 23}$Department of Physics and Astronomy, Tel Aviv University,
Tel Aviv 69978, Israel
\newline
$^{ 24}$International Centre for Elementary Particle Physics and
Department of Physics, University of Tokyo, Tokyo 113, and
Kobe University, Kobe 657, Japan
\newline
% ...
$^{ 29}$University of British Columbia, Department of Physics,
Vancouver BC V6T 1Z1, Canada
\newline
% ...
$^{  a}$ and at TRIUMF, Vancouver, Canada V6T 2A3
\newline
% ...
\newline
\clearpage
%
%
\section{Introduction}
\label{sec:intro}
 The measurement of the photon structure function \ft, and in particular
 of its evolution with the momentum transfer squared, \qsq,
 is a classic test of perturbative QCD~\cite{WIT-7701}.
 \ft is expected to increase only logarithmically with \qsq, therefore 
 the large range of \qsq values accessible at the \epem collider LEP,
 which presently extends from about 1~\gevsq to about 400~\gevsq 
 and which will increase in future, makes it an ideal place to study 
 this evolution.
 \par
 The results presented in this paper are based on new measurements of \ft 
 using data in the \qsq range from 6 to 100~\gevsq recorded by the 
 OPAL detector in 1996 at $\ssee=161-172$~\gev
 and on results of the analysis~\cite{OPALPR185,joergs}
 of data collected at \epem centre-of-mass energies \ssee around 
 the mass of the \zn (denoted by $\ssee=91$~\gev).
 This measurement is an extension of the analysis of \ft detailed
 in~\cite{OPALPR185} using basically the same methods to analyse
 the singly-tagged two-photon events at higher \ssee.
 \par
 For singly-tagged 
 events\footnote{The term singly-tagged denotes the situation where
 the electron, which radiates the virtual photon, 
 undergoes deep inelastic scattering and is seen (tagged) in the
 detector, whereas the other electron, which radiates the quasi-real photon,
 is unseen as it leaves the detector close to the beam direction. In this paper
 positrons are also referred to as electrons, and the electron and
 positron masses are neglected.}
 the process $\epem \rightarrow \epem + \mathit{hadrons}$ can be regarded 
 as deep inelastic scattering of an \epm on a quasi-real photon,
 where the flux of quasi-real photons can be calculated using the
 equivalent photon approximation~\cite{WEI-3401-WIL-3401-BUD-7501}.
 The cross-section for deep inelastic electron-photon scattering
 is expressed as~\cite{BER-8701}:
 \begin{equation}
  \frac{{\rm d}^2\sigma_{\rm e\gamma\rightarrow {\rm e X}}}{{\rm d}x{\rm d}Q^2}
 =\frac{2\pi\aemsq}{x\,Q^{4}}
  \left[\left( 1+(1-y)^2\right) \ftxq - y^{2} \flxq\right]
 \label{eqn:Xsect}
 \end{equation}
 where $\qsq=-q^2$ is the negative value of the four-momentum squared of the
 virtual photon. The usual dimensionless variables of deep inelastic
 scattering, $x$ and $y$, are defined as $x=\qsq/2(p\cdot q)$
 and $y=(p\cdot q)/(p\cdot k)$, and \aem is the fine structure constant.
 The symbols $p$, $q$ and $k$ denote the four-vectors of the 
 quasi-real photon, the virtual photon and the incoming electron
 respectively.
 The structure function \ft is related to the sum over the parton 
 densities of the photon (see e.g.~\cite{BER-8701}).
 In the kinematic region of low values of $y$ studied ($y^2\ll 1$) 
 the contribution of the term proportional to the longitudinal structure 
 function \flxq is negligible.
 \par
 The paper is organised as follows. After the description of the OPAL 
 detector in section~\ref{sec:detec} the data selection
 and background estimates are detailed in section~\ref{sec:selec}.
 The quality of the description of the observed 
 hadronic final state by the Monte Carlo models 
 is discussed in section~\ref{sec:descri}.
 The measurement of the evolution of \ft is outlined in 
 section~\ref{sec:f2extr}.
 The results are presented in section~\ref{sec:f2resu},
 followed by the conclusions given in section~\ref{sec:concl}.
%
%
\section{The OPAL detector}
\label{sec:detec}
 The OPAL detector is described in detail
 elsewhere~\cite{OPALPR021-ALL-9301-ALL-9401-AND-9401};
 only the subdetectors which are most relevant for this analysis, namely
 the electromagnetic calorimeters and the tracking devices, are
 detailed below.
 The OPAL detector has a uniform magnetic field of 0.435~T
 along the beam direction throughout the central tracking
 region, with electromagnetic and hadronic calorimetry and muon chambers
 outside the coil.
 \par
 In the OPAL right-handed coordinate system the $x$-axis points towards the
 centre of the LEP ring, the $y$-axis points upwards and the $z$-axis points in
 the direction of the electron beam.  The polar angle \thet and the
 azimuthal angle \ph are defined with respect to the $z$-axis
 and $x$-axis respectively.
 \par
 The small-angle silicon tungsten luminometer (SW)
 covers the region in \thet from 25 to 59~mrad.
 From 1996 onwards, the lower boundary of the acceptance has been increased
 to 32~mrad following the installation of a low angle shield to protect
 the central detector against possible synchrotron radiation.
 The SW consists of two finely segmented calorimeters placed around
 the beam pipe, one each side of the detector,
 at a distance of 2.4~m from the interaction point.
 Each calorimeter is composed of a stack of 19~silicon wafers
 interleaved with 18~tungsten plates placed perpendicular to 
 the beam axis. The sensitive area of each
 silicon wafer extends in radius from 62~to 142~mm from the beam axis and is
 segmented into 32~pads radially and 32~in azimuth around the beam.
 The radial position of electron showers in the calorimeter can be
 determined with a typical resolution of 0.06~mrad in \thet.
 The energy resolution for a beam energy electron of 86~\gev energy 
 is approximately 2.5~\gev. \par
 The clean acceptance of the forward detectors (FD) covers the \thet region 
 from 60 to 140~mrad at each end of the OPAL detector.
 The FD consists of cylindrical lead-scintillator calorimeters
 with a depth of 24 radiation lengths ($X_0$) divided azimuthally
 into 16 segments. The energy resolution for electromagnetic showers
 is $18\%/\sqrt{E}$, where $E$ is in \gev.
 An array of three planes of proportional tubes buried in the
 calorimeter at a depth of 4~$X_0$ provides a precise shower position
 measurement, with a typical resolution of 3--4~mm, corresponding to
 2.5~mrad in \thet, and no more than 3.5~mrad in \ph.
 \par
 Charged particles are detected by a silicon microvertex detector, a drift
 chamber vertex detector, and a jet chamber.
 Outside the jet chamber, but still in the magnetic field, lies
 a layer of drift chambers whose purpose is to improve the track
 reconstruction in the $z$-coordinate.
 The resolution of the transverse momentum for charged particles is \restr
 for $\act < 0.7$, where \pt is in \gev, and degrades for higher 
 values of \act.
 \par
 Both ends of the OPAL detector are equipped with
 electromagnetic endcap calorimeters covering the 
 range from 200 to 630~mrad in polar angle. They are homogeneous
 devices composed of arrays of lead-glass blocks of
 $9.2 \times 9.2$~cm$^2$ cross-section and typically 22 $X_0$
 in depth, giving good shower containment.
 In the central region, outside the solenoid, is the electromagnetic 
 barrel calorimeter of similar construction.
 \par
 The deep inelastic scattering events are triggered with high efficiency 
 by the large energy deposits of the scattered electron in the calorimeters 
 and by charged particle tracks seen in the tracking devices.
%
%
\section{Kinematics and data selection}
\label{sec:selec}
 The measurement of \ftxq requires the determination of $x$ and \qsq
 which are obtained from the four-vectors of the tagged
 electron and the hadronic final state as follows:
%
 \begin{equation}
  \qsq = 2\,\eb\,\etag\,(1- \cos\ttag)
 \end{equation}
 \begin{equation}
  x = \frac{\qsq}{\qsq+W^2+\psq}
 \label{eqn:Xcalc}
 \end{equation}
%
 Here \etag and \ttag are the energy and polar angle of the tagged
 electron, \eb is the beam energy, and $W$ the invariant mass of the hadronic
 final state.
 $\psq=-p^2$ is the negative value of the virtuality of the quasi-real photon.
 For this singly-tagged sample, an antitag condition on the second 
 electron is applied
 (see list of cuts below). This ensures that \psq is much smaller
 than \qsq and it is therefore neglected in the evaluation of $x$
 using Eq.~\ref{eqn:Xcalc}.
 \par
 The four-momentum of the hadronic system is calculated by summing over all
 charged particle tracks (assuming the pion mass) and calorimeter clusters,
 where quality criteria are applied to both the tracks and the clusters
 to ensure that they are well reconstructed.
 To avoid double counting of particle momenta, a matching algorithm
 for the association of a charged particle track with a cluster 
 is applied~\cite{MTPACK}.
 \par
 The analysis uses the data at $\ssee=161-172$~\gev
 with an integrated \epem luminosity of $18.1 \pm 0.1 $~\invpb, 
 as determined from small-angle Bhabha scattering events.
 The tagged electron is detected either in the SW detectors
 ($\qsq\approx 6-20\,\gevsq$)
 or in the FD detectors ($\qsq\approx 20-100\,\gevsq$).
 These two samples are subject to slightly different selection criteria 
 and are referred to as the SW and FD samples.
 Candidate events for the process \gghad are required to satisfy
 criteria for the tagged electron as well as for the hadronic final state,
 in addition to several technical cuts, to ensure good
 detector status and track quality.
 The event selection listed below is designed to have a high
 efficiency for signal events and to reject background events,
 which mainly stem from the process \ggtau.
 The requirements for the SW (FD) samples are:
%
 \begin{enumerate}
 \item A tagged electron candidate is required to produce
       a cluster in a SW (FD) detector with an energy of 
       $\etag\ge 0.775\,\eb\,(\etag\ge 0.60\,\eb)$ and a polar angle of
       $33 \le \ttag \le 55\,(60 \le \ttag \le 120)$~mrad with respect to
       either of the beam directions.
 \item The energy \ea of the most energetic cluster in the hemisphere
       opposite to the one which contains the tagged electron is restricted
       to $\ea\le 0.08\,\eb$ (antitag requirement).
 \item There must be at least three tracks originating from the hadronic
       final state, $\nch \ge 3$.
 \item The visible invariant mass \Wvis of the hadronic system,
       calculated as the mass of the four-momentum vector of the
       hadronic system as defined above,
       is required to be in the range $2.5 \le \Wvis \le 40$~\gev.
 \end{enumerate}
%
 The harder requirement on the tag energy applied to events of
 the SW sample reflects the much higher background from off-momentum 
 electrons closer to the beam direction. 
 An off-momentum electron is an electron 
 that was lost from the beam and scattered into the detector.
 In addition to the off-momentum electron, due to an interaction that 
 occurs close to the nominal 
 interaction point, charged particles are produced which 
 fulfill the requirements for the hadronic final state.
 Since the energy spectrum of the off-momentum electrons peaks
 around 0.5\eb this background is essentially eliminated
 by requiring $\etag\ge 0.775\,\eb$.\par
 With these cuts 879 and 414 events with average squared momentum
 transfers \qzm of approximately 11~\gevsq and 41~\gevsq,
 are selected in the SW and FD samples respectively.
 The accessible $x$ range for the two samples is $0.004 < x < 0.76$ and
 $0.012 < x <0.94$ respectively.
 The trigger efficiency is evaluated from the data to be above 98$\%$ for 
 the SW sample and essentially 100$\%$ for the FD sample.
 \par
 The background to the \gghad signal comes from
 events which contain a true or fake tagged electron and an apparent low-mass
 hadronic final state (compared to \ssee).
 The dominant source of this background is \gglep and particularly \ggtau.
 These processes were simulated with the Vermaseren 
 program~\cite{SMI-7701-BHA-7701-VER-7901-VER-8301}.
 The background contribution from \ggtau events is $ 36.0 \pm 1.8 $
 ($ 29.0 \pm 2.2 $) events and the \ggel process 
 contributes  $ 16.8 \pm 1.2 $ ($  9.1 \pm 0.9 $) events 
 to the SW (FD) samples respectively.
 Additional background sources like \znhad events, four-fermion events with
 \eeqq and \eell ($\ell = {\rm e}, \mu, \tau$)
 final states, and \zntau events, were studied.
 The total contribution of these processes to the background 
 is of the order of 1 and 3 events in the two samples, respectively, and
 consequently they are neglected.
 \par
 Events with a scattered electron observed in the electromagnetic endcap 
 calorimeter allow higher values of \qsq to be reached than in the SW and 
 FD samples. With the current level of integrated luminosity 
 collected at $\ssee=161-172$~\gev too few such events have been observed
 to allow a detailed analysis to be performed.
%
%
\section{Description of the hadronic final state}
\label{sec:descri}
 The measurement of \ft
 requires the determination of $W$ from the hadronic final state.
 Because of the finite detector resolution and the incomplete
 angular coverage, especially in the forward region,
 the correlation between visible hadronic mass, \Wvis, and $W$ 
 depends critically on the modelling of the hadronic final state, which is
 correlated with the observed hadronic energy flow.
 Therefore a detailed comparison of the observed hadronic final state
 and the predictions from the various Monte Carlo models is needed.
 The results of this study, which follows closely the one described
 in~\cite{OPALPR185}, are summarised in this section.
 \par
 The set of Monte Carlo generators used to simulate the signal events
 differs only slightly from that used in the previous 
 analysis~\cite{OPALPR185}.
 HERWIG~\cite{HERWIG} version 5.9 is used instead of 5.8d with the 
 charm quark mass altered from 1.8~\gev to 1.55~\gev.
 All events are generated without the simulation of the soft underlying
 event~\cite{OPALPR185}.
 The F2GEN~\cite{BUI-9401} generator is used to simulate events
 based on \ft for four flavours (u,d,s,c) with $\mc = 1.5$~\gev,
 rather than with three flavours and with the charm contribution
 added using the Vermaseren program, as in~\cite{OPALPR185}.
 The PYTHIA~\cite{SJO-9401-SJO-9301} Monte Carlo remains the same
 as in~\cite{OPALPR185}. 
 The integrated luminosity of each Monte Carlo sample corresponds 
 to approximately ten times the integrated luminosity of the data. 
 The Monte Carlo generation was based either on the leading order
 (LO) GRV~\cite{GLU-9201-GLU-9202} or the LO SaS1D~\cite{SCH-9501}
 parton density parametrisations.
 All Monte Carlo events are passed through the OPAL detector
 simulation program~\cite{ALL-9201} and the same reconstruction and
 analysis chain as the real data events.
 \par
 For the comparison of the observed hadronic final states
 to the Monte Carlo predictions
 a particle is defined as a track or a cluster resulting
 from the matching algorithm~\cite{MTPACK}.
 In~\cite{OPALPR185} it was shown that none of the models used
 accurately described the angular distribution and the transverse
 energy of the hadronic final state.  
 These discrepancies persist at higher centre-of-mass energies.
 The accessible \qsq region for the SW sample is similar to that 
 of the low-\qsq sample of the analysis at \ssee = 91~\gev~\cite{OPALPR185},
 where the scattered electron was observed in the FD.
 \par
\begin{figure}
\begin{center}
\mbox{\epsfig{file=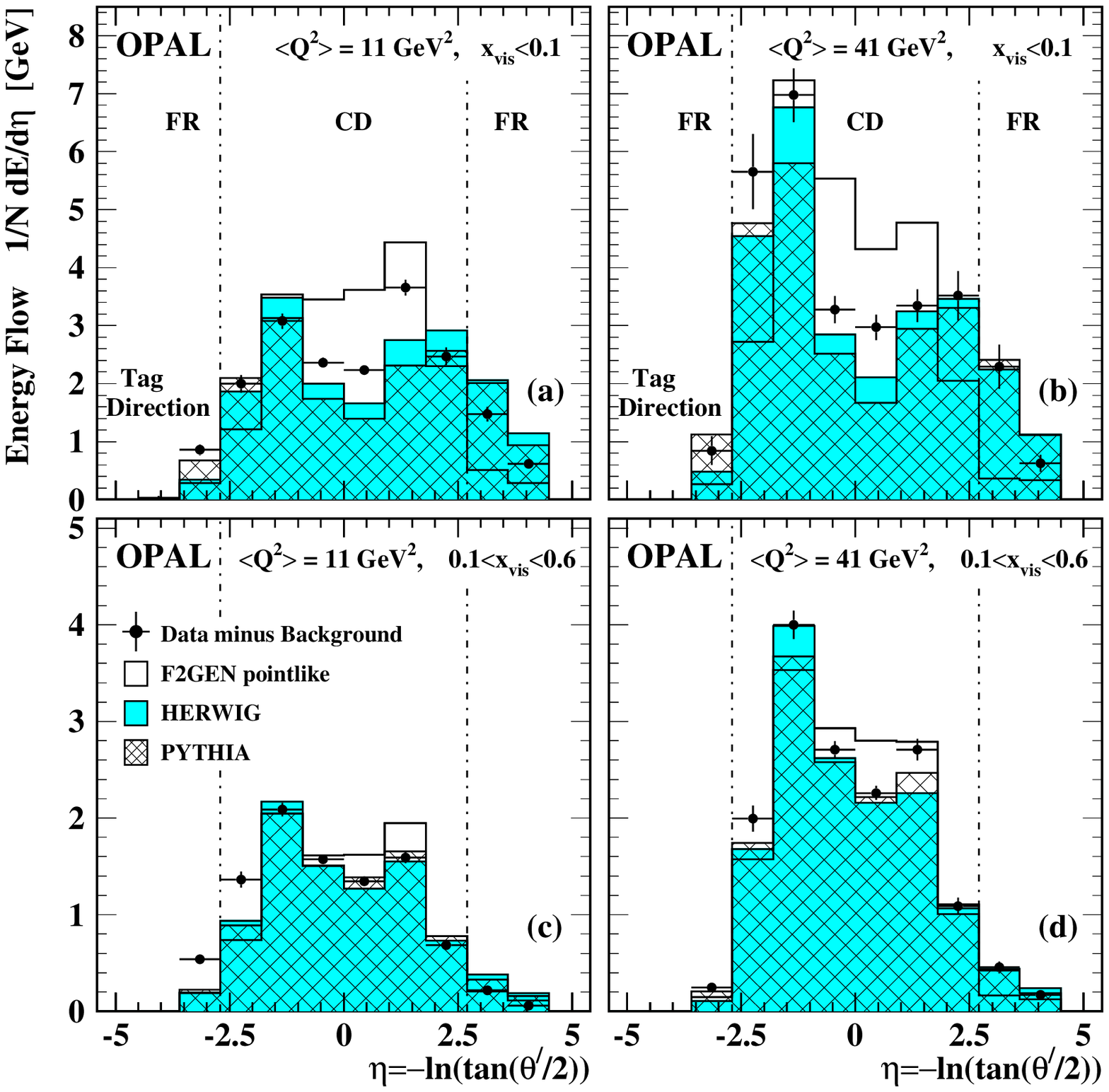,height=15cm}}
\caption{\label{fig:pr207_01}
 The hadronic energy flow per event as a function of pseudorapidity $\eta$
 for the data taken at $\protect\ssee=161-172$~\gev
 and various signal Monte Carlo samples. The energy flow is shown
 for two ranges of \xvis for the SW and FD samples.
 The errors shown are statistical only. The tagged electron is always
 at negative pseudorapidity and is not shown.
 The different regions in rapidity indicated denote the
 forward region (FR) which contains the SW and FD detectors
 and the central detector (CD).
 }
\end{center}
\end{figure}

 Figure~\ref{fig:pr207_01} shows the hadronic energy flow
 per event, \flow, as a function of pseudorapidity, 
 $\eta=-\ln(\tan(\theta'/2))$, 
 for the SW sample and the FD sample for two bins in \xvis. 
 Only statistical errors are shown. The tagged electron is
 always at negative pseudorapidity and is not shown.
 In the region $0.1 < \xvis < 0.6$
 the agreement between data and the various models is satisfactory, with
 the exception of the region $-4 < \eta < -2$ for the SW sample where
 all Monte Carlos are low compared to the data. 
 This effect comes from leakage into the FD detector and 
 is almost entirely eliminated if the requirement for the 
 maximum \ttag for the SW sample is restricted to smaller angles. 
 This is taken into account in the evaluation of the systematic error on \ft.
 Significant differences are observed for $\xvis < 0.1$.
 In the remnant direction at positive $\eta$, both HERWIG and PYTHIA
 tend to overestimate the energy deposited in the forward region of the
 detector. In the central region, however, they underestimate the 
 transverse energy.
 In contrast, the pointlike events generated with the F2GEN model 
 lie higher than the data in the central region and 
 lower in the forward region.
 The number of events with $\xvis > 0.6$ is insufficient 
 to allow firm conclusions to be drawn.
 \par
 The minimum tag energy and the antitag requirement
 for the FD sample of this analysis are slightly
 different from the ones used for the low-\qsq sample in~\cite{OPALPR185}.
 This results in a different shape of the energy flow distributions
 and a slightly improved quality of the data description by the Monte
 Carlo models.
 \par
 In summary, in the intermediate range of \xvis a good overall agreement
 between the data and the Monte Carlo models is found for both samples.
 At $\xvis < 0.1$ the flow of hadronic energy into the central region of 
 the detector is underestimated by the HERWIG and PYTHIA Monte Carlo models,
 and overestimated by the F2GEN model.
 Improved  Monte Carlo models would be therefore required in order
 to reduce the model-dependent systematic errors on the measurement 
 of \ft in this region.
%
%
\section{Determination of \boldmath{\ft}}
\label{sec:f2extr}
 To obtain the photon structure function \ft 
 in bins of \qsq from the measured \xvis distribution the method of
 regularised unfolding is employed~\cite{BLO-8401-BLO-9601}.
 In this method, a reference Monte Carlo is used to 
 obtain the detector efficiency and resolution and the regularisation 
 reduces the effect of inherent oscillating fluctuations of the unfolded 
 structure function. See~\cite{OPALPR185} for more details of this procedure.
 To unfold \ft from the data HERWIG, with the \ft structure 
 function based on the GRV 
 parametrisation, is chosen as the reference Monte Carlo.
 This choice is motivated by the fact that HERWIG gives a 
 slightly better description of the hadronic final state than 
 PYTHIA, (see~\cite{OPALPR185} and Fig.~\ref{fig:pr207_01}).
 The reference Monte Carlo is used to determine the
 central values and most of the systematic uncertainties.
 The PYTHIA and the F2GEN models are used to estimate the model
 dependence of the result.
 \par
\begin{figure}
\begin{center}
\mbox{\epsfig{file=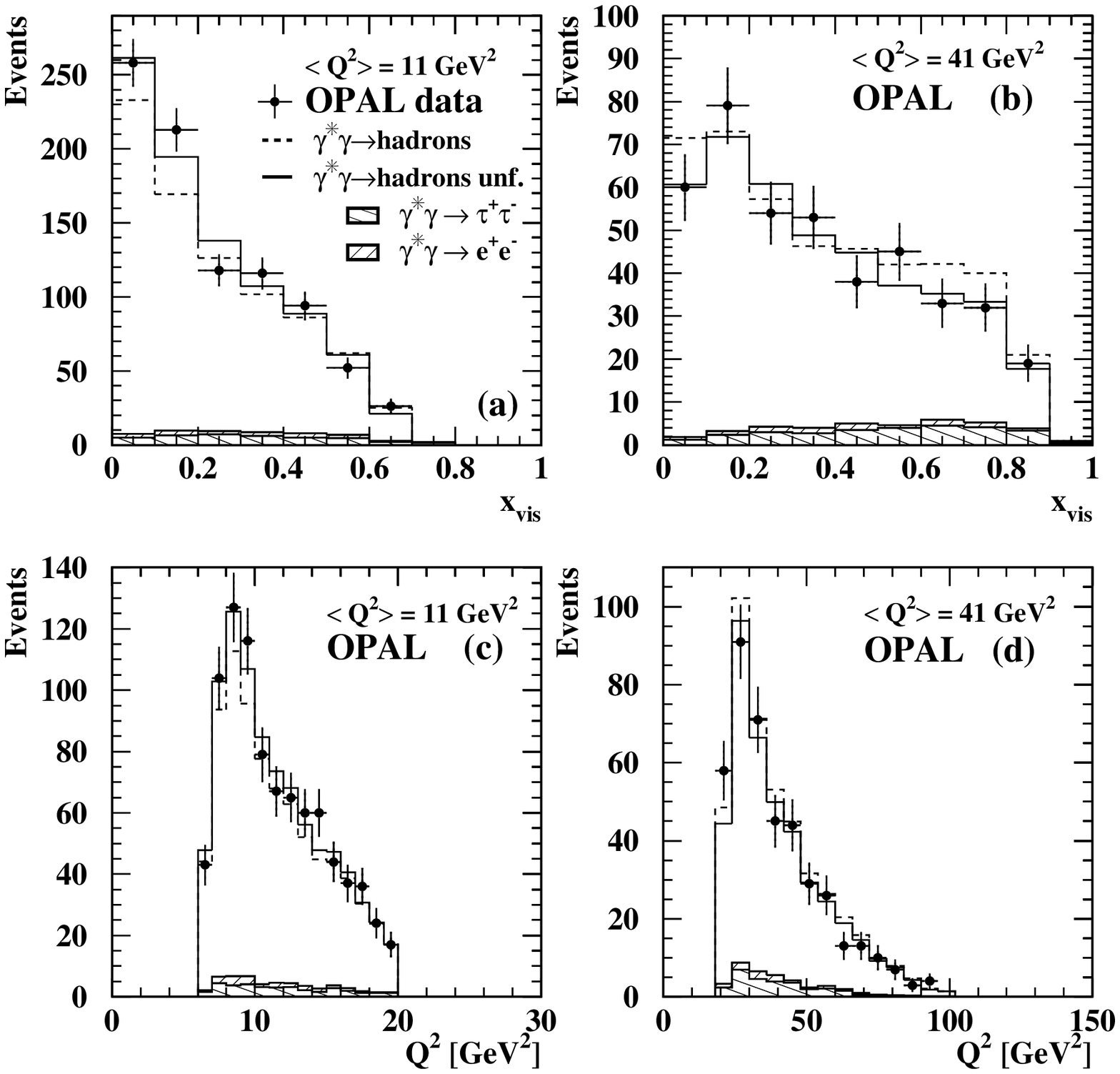,height=17cm}}
\caption{\label{fig:pr207_02}
 The \xvis, (a) and (b), and \qsq, (c) and (d), distribution of 
 the SW and FD samples for the data taken at $\protect\ssee=161-172$~\gev.
 The \textit{dashed} histogram, \gghad,
 shows the events from the HERWIG Monte Carlo,
 using the GRV parametrisation and the standard cuts (the reference sample),
 with the background added, before the unfolding; the
 \textit{solid} histogram, \gghad \textit {unf.}, 
 shows the same quantity, but reweighted based 
 on the result of the unfolding. The background events from the reactions 
 \ggtau and \ggel are also shown separately at the bottom of the figure.
 All errors shown are statistical only.
 }
\end{center}
\end{figure}

 The data at $\ssee=161-172$~\gev are subdivided into four \qsq ranges, two 
 ranges for the SW sample and two for the FD sample. 
 The ranges in \qsq are 6--11, 11--20, 20--40 and 40--100~\gevsq
 with average squared momentum transfers
 of \qzm = 9, 14.5, 30 and 59~\gevsq respectively.
 The \xvis distributions of the SW and FD samples are shown in
 Figs.~\ref{fig:pr207_02}a and~\ref{fig:pr207_02}b.
 The distributions of the reference Monte Carlo with background
 added are shown by the dashed histograms. In addition, the background
 events are shown separately at the bottom of the figure. The number of 
 Monte Carlo events is the absolute prediction for the data luminosity.
 The solid histogram represents the reweighted distribution of 
 the signal Monte Carlo with background added
 after the unfolding has been performed~\cite{OPALPR185}.
 From the figures one observes
 that the mean \xvis increases with increasing \qsq,
 and that the \xvis distribution of the data is well represented by the
 sum of the reweighted signal Monte Carlo and background Monte Carlo samples
 after unfolding.
 \par
 The weight factors for each individual event, obtained from the unfolding, 
 can be used to construct reweighted Monte Carlo distributions of 
 different variables. 
 The reweighted distributions are the Monte Carlo predictions 
 based on the structure function unfolded from the data.
 Any reweighting based on the generated $x$ distribution will
 change the shape of other measurable variables besides \xvis.
 As all variables are differently correlated with \xvis 
 the comparison of their distributions with the data gives 
 an important
 check of the transformation, as described by the Monte Carlo simulation,
 between the partonic distributions and the measurable distributions.
 As an example Figs.~\ref{fig:pr207_02}c and ~\ref{fig:pr207_02}d 
 show the \qsq distributions of the SW and FD samples,
 applying the cuts listed in Sect.~\ref{sec:selec}.
 There is good agreement in the shape of the distributions.
 The distributions from both Monte Carlo samples differ slightly 
 in normalisation from the distributions observed in the data,
 and the agreement improves after the reweighting based on
 the unfolding of the \xvis distribution.
 A similar behaviour is found for other observables,
 both for quantities measured from the electron, like \etag and \ttag,
 and for distributions obtained from the hadronic final state, like \nch and
 \Wvis.
%
\section{Results}
\label{sec:f2resu}
\begin{figure}
\begin{center}
\mbox{\epsfig{file=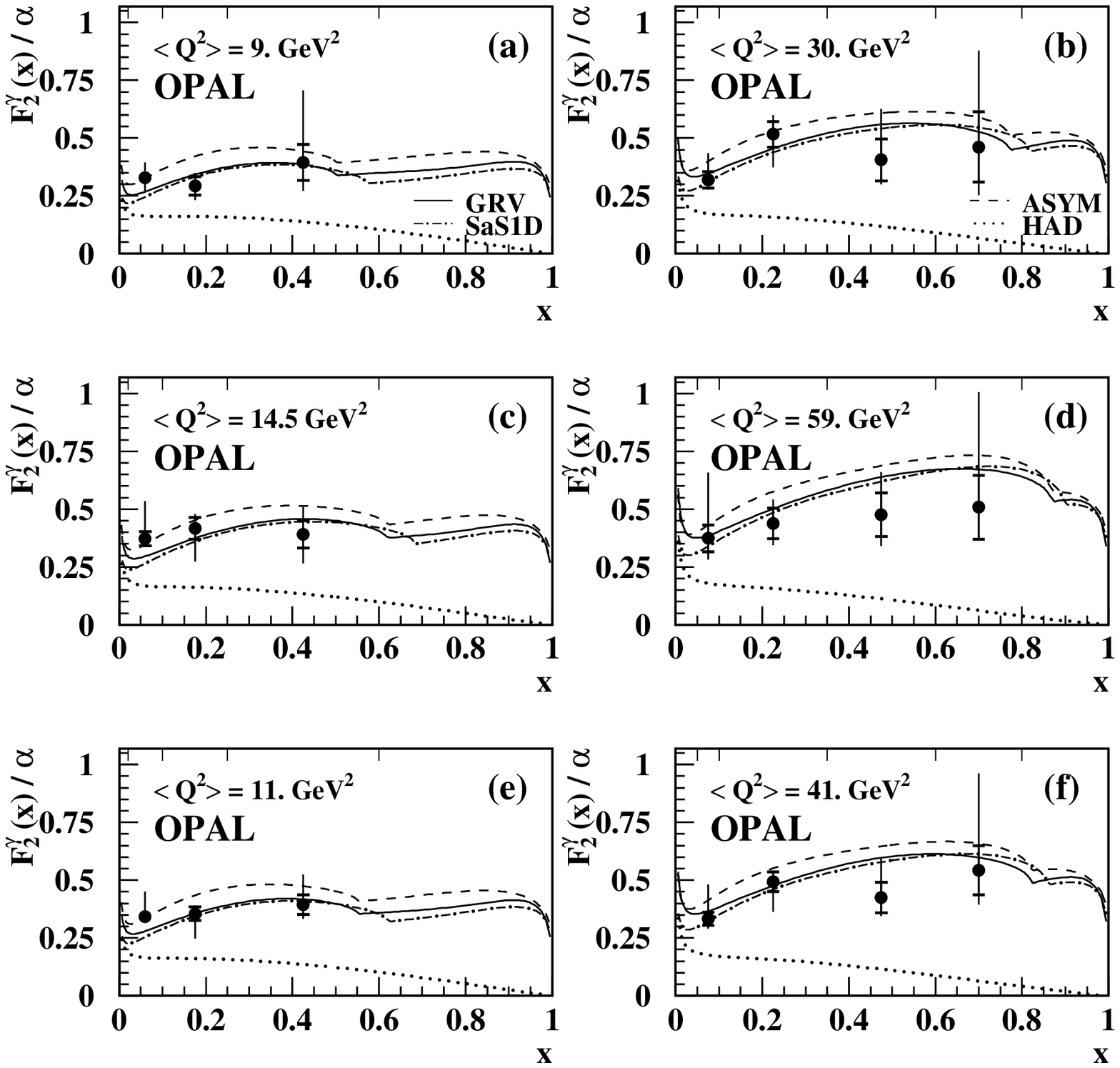,height=17cm}}
\caption{\label{fig:pr207_03}
 The measurement of \ft for the data taken at $\protect\ssee=161-172$~\gev.
 The structure function \ft is measured for four active flavours 
 in four bins in \qsq with mean values of
 (a) \qzm =  9~\gevsq, (b) \qzm = 30~\gevsq, (c) \qzm = 14.5~\gevsq,
 and (d) \qzm = 59~\gevsq.
 In (e) the measurement for the combined data sets of (a) and (c),
 and in (f) the measurement for the combined data sets of
 (b) and (d) is shown.
 The \textit{points} show the measured \ft. The bin sizes are indicated by
 the \textit{vertical lines} at the top of the figure.
 The \textit{solid} line represents the \ft derived from the GRV
 parametrisation and the \textit{dot-dashed} line denotes
 the \ft derived from the SaS1D parametrisation, both using the
 Bethe-Heitler contribution to \ft for massive charm quarks.
 The charm quark mass \mc is taken to be 1.3~\gev and 1.5~\gev in the case of
 SaS1D and GRV, respectively.
 The \textit{dotted} line (HAD) represents the hadronic component and 
 the \textit{dashed} line (ASYM) the full augmented asymptotic \ft, as
 described in the text.
 }
\end{center}
\end{figure}

 The unfolded \ft measurements for the data taken at $\ssee=161-172$~\gev
 are shown in Figs.~\ref{fig:pr207_03}a$-$\ref{fig:pr207_03}f, and are 
 listed in Table~\ref{tab:resxq}.
 The value of \ftn is shown at the centre of the $x$ bin.
 The bin sizes are indicated by the vertical lines at the top of the figure.
 The error bars show both the statistical error alone and the full
 error, given by the quadratic sum of statistical and systematic errors.
 The central values and statistical errors of the \ft measurements
 are estimated by using the reference unfolding.
 The \ft values presented here are not corrected for the fact 
 that \psq is not strictly equal to zero~\cite{OPALPR185}.
 \par
%
\renewcommand{\arraystretch}{1.20}
\begin{table}[htb]
\begin{center}
\begin{tabular}{cccc}
 \hline
 \qsq range  $[\gevsq]$ &  \qzm $[\gevsq]$ & $\Delta x$ & \ftn  \\
 \hline
6$-$11   & 9    & $0.020 - 0.100$ & \Z{0.33}{0.03}{0.06}{0.06}\\
         &      & $0.100 - 0.250$ & \Z{0.29}{0.04}{0.04}{0.05}\\
         &      & $0.250 - 0.600$ & \Z{0.39}{0.08}{0.30}{0.10}\\
 \hline
11$-$20  & 14.5 & $0.020 - 0.100$ & \Z{0.37}{0.03}{0.16}{0.01}\\
         &      & $0.100 - 0.250$ & \Z{0.42}{0.05}{0.04}{0.14}\\
         &      & $0.250 - 0.600$ & \Z{0.39}{0.06}{0.10}{0.11}\\
 \hline
20$-$40  & 30   & $0.050 - 0.100$ & \Z{0.32}{0.04}{0.11}{0.02}\\
         &      & $0.100 - 0.350$ & \Z{0.52}{0.05}{0.06}{0.13}\\
         &      & $0.350 - 0.600$ & \Z{0.41}{0.09}{0.20}{0.05}\\
         &      & $0.600 - 0.800$ & \Z{0.46}{0.15}{0.39}{0.14}\\
 \hline
40$-$100 & 59   & $0.050 - 0.100$ & \Z{0.37}{0.06}{0.28}{0.07}\\
         &      & $0.100 - 0.350$ & \Z{0.44}{0.07}{0.08}{0.07}\\
         &      & $0.350 - 0.600$ & \Z{0.48}{0.09}{0.16}{0.10}\\
         &      & $0.600 - 0.800$ & \Z{0.51}{0.14}{0.48}{0.02}\\
 \hline
 \hline
6$-$20   & 11   & $0.020 - 0.100$ & \Z{0.34}{0.02}{0.11}{0.02}\\
         &      & $0.100 - 0.250$ & \Z{0.36}{0.03}{0.02}{0.10}\\
         &      & $0.250 - 0.600$ & \Z{0.39}{0.04}{0.12}{0.04}\\
 \hline
20$-$100 & 41   & $0.050 - 0.100$ & \Z{0.33}{0.03}{0.15}{0.03}\\
         &      & $0.100 - 0.350$ & \Z{0.49}{0.04}{0.02}{0.12}\\
         &      & $0.350 - 0.600$ & \Z{0.43}{0.07}{0.15}{0.05}\\
         &      & $0.600 - 0.800$ & \Z{0.54}{0.11}{0.41}{0.11}\\
\hline
\end{tabular}
\caption{Results for \ft as a function of $x$ for four active flavours
         in bins of \qsq for the data taken at $\protect\ssee=161-172$~\gev.
         The first errors are statistical and the second systematic.
         The SW sample with $\qzm = 11$~\gevsq consists of the data from 
         $\qzm = 9$ and 
         14.5~\gevsq and the FD sample with $\qzm = 41$~\gevsq consists 
         of the data from $\qzm = 30$ and 59~\gevsq.%
         }
\label{tab:resxq}
\end{center}\end{table}
%
 The estimation of the systematic error includes three parts~\cite{OPALPR185}:
 the variation of the compositions of signal and background events
 in the sample, the use of different \ft structure functions assumed in the
 Monte Carlo samples, and the different modelling of the formation
 of the hadronic final state.
 To allow for a different composition of signal and background events,
 the event selection cuts are varied.
 The choice of the cut variations reflects the different population
 of signal events in the four \qsq ranges, in terms of the scattering angle
 of the electron and \Wvis, as well as the different behaviour of the
 background events.
 In each case the unfolding is carried out using the reference Monte Carlo 
 model where only one cut is varied from the standard set.
 To study the uncertainty due to the structure functions assumed in the
 Monte Carlo samples, for the SW sample, the unfolding is done
 using the HERWIG generator, the standard set of cuts and the
 SaS1D parton density parametrisations.
 The effect of the different modelling of the formation of the hadronic
 final state is studied by repeating the unfolding using the PYTHIA
 Monte Carlo model with the SaS1D parametrisation,
 and the F2GEN Monte Carlo model with the GRV parametrisation.
 The systematic error assigned to the result, shown in Table~\ref{tab:resxq},
 is taken as the maximum deviation of any unfolding result
 from the central values, which in some cases leads to rather 
 asymmetric errors, particularly when the HERWIG Monte Carlo model is
 replaced by PYTHIA or F2GEN.
 The systematic error is dominated by the model uncertainties.
 \par
 The measured \ft as function of $x$ is almost flat within the region studied 
 and the absolute normalisation of \ft is well described by
 various predictions.
 Shown in Fig.~\ref{fig:pr207_03} are the \ft predictions 
 for the leading order GRV and the SaS1D parton density parametrisations
 evaluated at the corresponding values of \qzm. 
 In each case the expected contribution from massive charm quarks is added.
 The fact that the charm threshold depends on $x$ and varies with \qsq
 causes the change in shape of the predictions, see Eq.~\ref{eqn:Xcalc} 
 for $W=2\mc$.
 Both parametrisations successfully describe the measured \ft as a 
 function of $x$ in all \qsq ranges.
 \par
 Figure~\ref{fig:pr207_03} also shows an augmented asymptotic 
 prediction for \ft.
 The contribution to \ft from the three light flavours is 
 approximated by Witten's leading order asymptotic form~\cite{WIT-7701},
 using the parametrisation given in~\cite{GOS-9201}. 
 This has been augmented by adding a charm contribution 
 evaluated from the Bethe-Heitler formula~\cite{WIT-7601}, and an estimate of 
 the hadronic part of the photon structure function, which essentially 
 corresponds to the hadronic part of the GRV (LO) parametrisation of \ft.
 The hadronic contribution is derived from the structure 
 function parametrisation of the pion~\cite{GLU-9203} using VDM arguments
 as detailed in~\cite{GLU-8401} 
 and evolved to the corresponding values of \qzm.
 The components are evaluated in leading order using \lamt = 0.232 \gev.
 It is known~\cite{GLU-8301,GLU-8401} 
 that the asymptotic solution, which is obtained by neglecting 
 terms in the evolution equations that vanish for 
 $\qsq\rightarrow\infty$, has deficits in the region of low $x$, 
 because the remaining terms create divergences in the solution
 which do not occur in the solution of the full evolution 
 equations~\cite{GLU-8401}. 
 However, the asymptotic solution has the appealing feature that 
 it is calculable in QCD, even at higher order~\cite{BAR-7901} and 
 for medium $x$ and with increasing \qsq it should be more reliable.
 In addition at high $x$ and \qsq the hadronic contribution is expected 
 to be small. 
 In the region of medium $x$ values studied here this asymptotic prediction 
 in general lies higher than the GRV and SaS predictions but it is still 
 in agreement with the data.
 The importance of the hadronic contribution to \ft, which is shown separately 
 at the bottom of the figure, decreases with increasing $x$ and \qsq, 
 and it accounts for only 15$\%$ of \ft at $\qsq = 59$~\gevsq and $x = 0.5$.
 The asymptotic solution increases with decreasing \lamt.
 For $\qsq = 59$~\gevsq and $x = 0.5$ the change in \ft is $+24\%$ 
 and $-16\%$ if \lamt is changed from $\lamt=0.232$~\gev
 to 0.1~\gev and 0.4~\gev respectively.
 \par
\begin{figure}
\begin{center}
\mbox{\epsfig{file=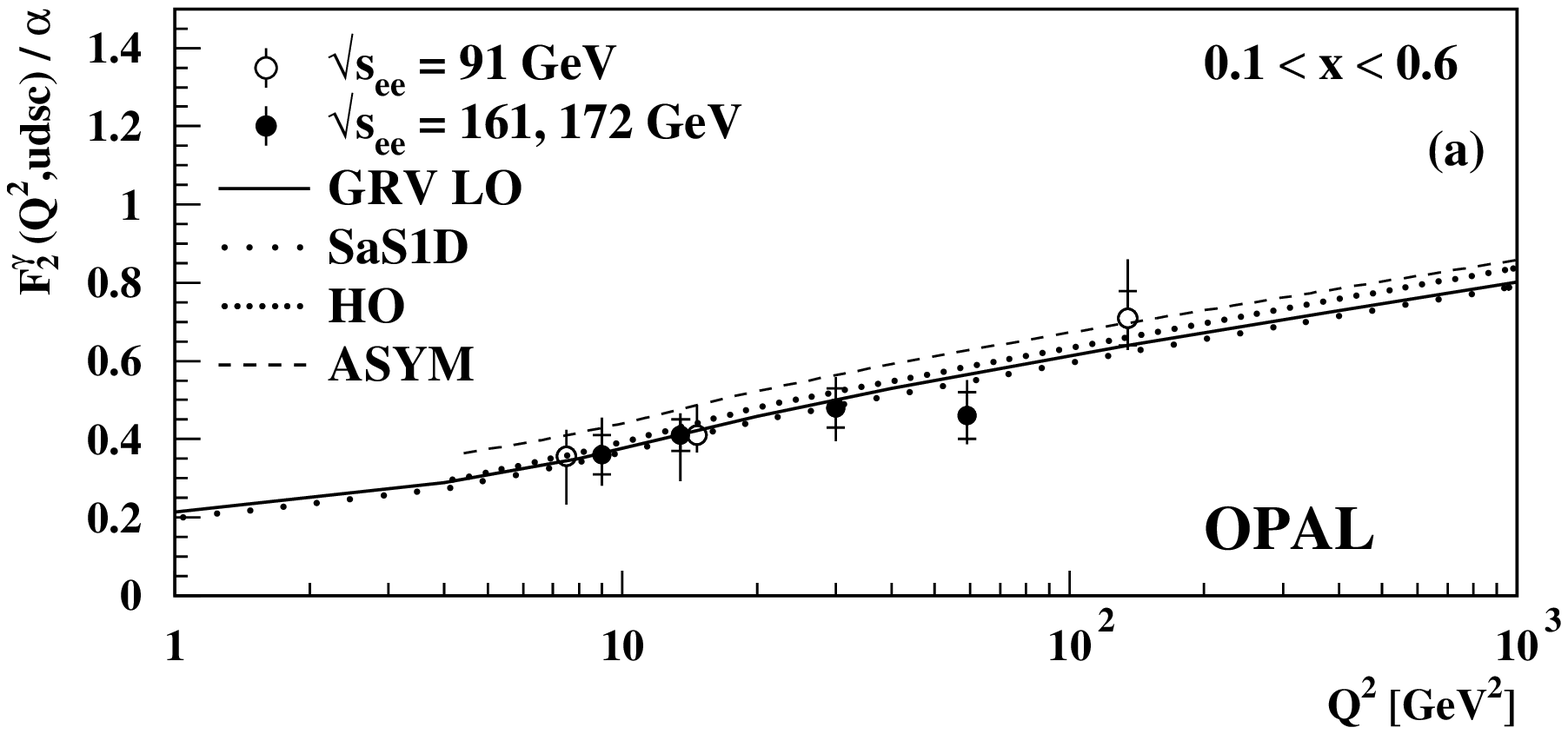,width=15.cm}}
\mbox{\epsfig{file=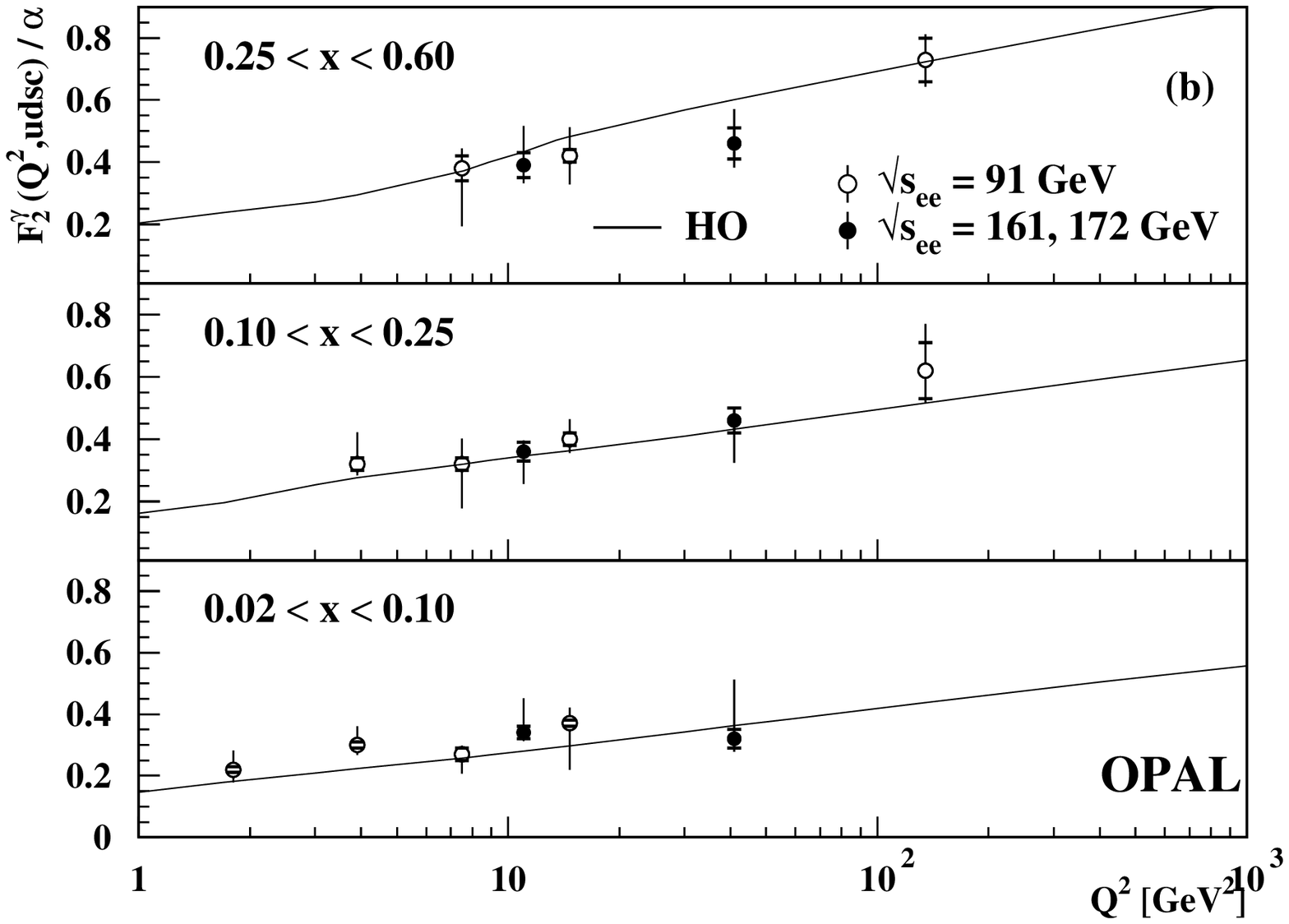,width=15.cm}}
\caption{\label{fig:pr207_04}
 The measurement of \ft for four active flavours as a function of \qsq,
 (a) for the range $0.1<x<0.6$, and 
 (b) subdivided into $0.02<x<0.10$, $0.10<x<0.25$ and $0.25<x<0.60$.
 In addition shown in (a) are the \ft of the GRV (LO) 
 and the SaS1D (LO) parametrisation, the \ft of the asymptotic prediction
 (ASYM) and the result of a higher order calculation (HO),
 where the last two predictions are only shown for $\qsq >4$~\gevsq.
 In (b) the data are only compared to the HO prediction.
 In both figures the errors are statistical and systematic. In some of the 
 cases the statistical errors are not visible because they are smaller 
 than the size of the symbols.
 }
\end{center}
\end{figure}

 As in the previous analysis, \ftn is also determined in bins of \qsq 
 in the region $0.1<x<0.6$.
 The new results, together with the result of the previous 
 analysis~\cite{OPALPR185}, are shown as a function of \qsq in
 Fig.~\ref{fig:pr207_04}a and listed in Table~\ref{tab:resmea}.
 For the results based on the data taken at $\ssee=161-172$~\gev 
 the systematic error is evaluated as described above.
 The results of the two analyses at very different centre-of-mass energies,
 using different detector components to detect the scattered electron
 for similar \qsq values, are found to be consistent.
 An increase of \ftn with \qsq is observed in the data, in agreement with 
 the QCD predictions and with results obtained by other 
 experiments~\cite{TPC-8701-TPC-8702-DEL-9601-JAD-8401-TOP-9402}.
 The data are compared
 to the LO predictions of the GRV and the SaS1D parametrisations, both
 including the contribution to \ft from massive charm quarks, and
 to a higher order (HO) calculation~\cite{LAE-96xx}
 based on the HO GRV parametrisation for three light quarks,
 complemented by the contribution of charm
 quarks to \ft based on the HO calculation using massive charm quarks 
 of~\cite{LAE-9401}, as described in more detail in~\cite{OPALPR185}.
 The differences between the three predictions are small compared to the 
 experimental errors, and all predictions nicely agree with the data.
 \par
 In addition the data are compared to the asymptotic prediction as
 detailed above. This approximation lies higher than the data 
 at low \qsq and approaches the data at the highest \qsq reached.
 \par
%
\renewcommand{\arraystretch}{1.20}
\begin{table}[htb]
\begin{center}\begin{tabular}{ccc}
\hline
    \qzm    &   \ftn    & Ref.\\
\hline
  7.5 & \Z{0.36}{0.02}{0.06}{0.12} & \protect\cite{OPALPR185}\\
 14.7 & \Z{0.41}{0.01}{0.08}{0.04} & \protect\cite{OPALPR185}\\
  135 & \Z{0.71}{0.07}{0.14}{0.05} & \protect\cite{OPALPR185}\\
\hline
\hline
   9  & \Z{0.36}{0.05}{0.08}{0.06}&\\
 14.5 & \Z{0.41}{0.04}{0.04}{0.11}&\\
   30 & \Z{0.48}{0.05}{0.06}{0.07}&\\
   59 & \Z{0.46}{0.06}{0.07}{0.04}&\\
 \hline
   11 & \Z{0.38}{0.03}{0.06}{0.03}&\\
   41 & \Z{0.47}{0.04}{0.06}{0.03}&\\
\hline
\end{tabular}
\caption{Results for \ft for four active flavours in bins of \qsq 
         for $0.10 < x < 0.60$.
         The first errors are statistical and the second systematic.
         The bins in \qsq for the analysis 
         of the data taken at $\protect\ssee=161-172$~\gev are defined in 
         Table\protect~\ref{tab:resxq}.
         The SW sample with $\qzm = 11$~\gevsq consists of the data from 
         $\qzm = 9$ and 
         14.5~\gevsq and the FD sample with $\qzm = 41$~\gevsq consists 
         of the data from $\qzm = 30$ and 59~\gevsq.
         The combined results at \qzm = 11 and 
         41~\gevsq are shown for completeness and are 
         not used in the fit of the slope of 
         \ft, Fig.~\protect\ref{fig:pr207_04}a.
         }
\label{tab:resmea}
\end{center}\end{table}
%
 The evolution of \ft with \qsq is measured by fitting
 a linear function of the form $a + b\,\ln (\qsq/\gevsq)$
 to the four data points obtained at $\ssee=161-172$~\gev.
 Here $a$ and $b$ are parameters which do not depend on $x$.
 The result is
 $$\ftqn=(0.24 \pm 0.11^{+0.06}_{-0.18}) + 
         (0.06 \pm 0.04^{+0.05}_{-0.02})\,\ln\,(\qsq/\gevsq) \, .$$
 The central values and statistical errors of the two parameters are obtained 
 by a fit to the central values of the individual points listed 
 in Table~\ref{tab:resmea} using their statistical errors.
 The quality of the fit is satisfactory with $\chidof = 0.33$ for the 
 central value. 
 The systematic errors are obtained from the maximum deviation from the central
 values to either side observed when repeating the fit for 
 all variations of cuts, structure functions assumed in the Monte Carlo 
 and Monte Carlo models as discussed above. 
 This procedure automatically takes into account 
 the correlation of the systematic uncertainties between the individual points.
 \par
 This measurement is consistent with the result
 of the previous analysis~\cite{OPALPR185} which used the data sample 
 at $\ssee=91$~\gev in the \qsq range 7.5$-$135~\gevsq, 
 and which was based on a slightly different fitting procedure.
 A combined fit to the OPAL data at $\ssee = 91$~\gev and $161-172$~\gev
 in the \qsq range of 7.5$-$135~\gevsq yields:
 $$\ftqn=(0.16 \pm 0.05^{+0.17}_{-0.16}) +
         (0.10 \pm 0.02^{+0.05}_{-0.02})\,\ln\,(\qsq/\gevsq)$$
 with $\chidof = 0.77$ for the central value. 
 The slope \slop is significantly different from zero and the error is
 somewhat reduced compared to~\cite{OPALPR185}
 due to the inclusion of the data at higher centre-of-mass energies.
 \par
 The QCD prediction for the scaling violation behaviour of \ft is
 different from the proton case.
 With increasing \qsq, the proton structure function, \ftp, increases at 
 small $x$ and decreases at large $x$, ($x \gtrsim 0.1$).
 In contrast, due to the pointlike coupling of the photon to quarks,
 the photon structure function \ft is expected to increase
 with \qsq for all values of $x$, and
 the size of the scaling violation is expected to depend on $x$. 
 To examine whether the data exhibit the predicted variation in 
 \slop, the \qsq range 1.86$-$135~\gevsq is analysed
 using common $x$ ranges, as listed in Table~\ref{tab:resdif}.
 The published results are rebinned 
 in $x$ but not reanalysed, i.e. the systematic error is evaluated
 as described in~\cite{OPALPR185,joergs}.
 \par
%
\renewcommand{\arraystretch}{1.20}
\begin{table}[htb]
\begin{center}\begin{tabular}{ccccc}
\hline
 $\Delta x$ & $0.02-0.10$ & $0.10-0.25$ & $0.25-0.60$ &\\
    \qzm    &        \ftn &        \ftn &        \ftn &Ref.\\
\hline
  1.86&\Z{0.22}{0.01}{0.06}{0.04}  & $-$
      & $-$                        &\protect\cite{joergs}\\
  3.76&\Z{0.30}{0.01}{0.06}{0.03}  & \Z{0.32}{0.02}{0.10}{0.03}
      & $-$                        &\protect\cite{joergs}\\
  7.5 & \Z{0.27}{0.02}{0.02}{0.06} & \Z{0.32}{0.02}{0.08}{0.14}
      & \Z{0.38}{0.04}{0.05}{0.18} & \protect\cite{OPALPR185}\\
 14.7 & \Z{0.37}{0.01}{0.05}{0.15} & \Z{0.40}{0.02}{0.06}{0.04}
      & \Z{0.42}{0.02}{0.09}{0.09} & \protect\cite{OPALPR185}\\
  135 &           $-$              & \Z{0.62}{0.09}{0.12}{0.05}
      & \Z{0.73}{0.07}{0.04}{0.05} & \protect\cite{OPALPR185}\\
\hline
\hline
   11 & \Z{0.34}{0.02}{0.11}{0.02} & \Z{0.36}{0.03}{0.02}{0.10}
      & \Z{0.39}{0.04}{0.12}{0.04} &\\
   41 & \Z{0.32}{0.03}{0.19}{0.03} & \Z{0.46}{0.04}{0.02}{0.13}
      & \Z{0.46}{0.05}{0.10}{0.06} &\\
\hline
\end{tabular}
\caption{Results for \ft for four active flavours in bins of \qsq and $x$.
         The first errors are statistical and the second systematic.
         The bins in \qsq for the analysis 
         of the data taken at $\protect\ssee=161-172$~\gev are defined in 
         Table\protect~\ref{tab:resxq}.
         The results from~\protect\cite{OPALPR185,joergs}
         are rebinned to match the same ranges in $x$.
         For statistical reasons only the combined
         samples with $\qzm = 11$ and 41~\gevsq 
         are used for the subdivision into several
         bins in $x$, Fig.~\protect\ref{fig:pr207_04}b.
         }
\label{tab:resdif}
\end{center}\end{table}
%
 Figure~\ref{fig:pr207_04}b shows the measurement in comparison 
 to the HO calculation. The points of inflection of \ft for \qsq 
 below 15~\gevsq are caused by the charm threshold.
 The predictions are in agreement with the observed evolution of 
 \ft with \qsq in all ranges of $x$.
 In order to observe experimentally the variation of \slop with $x$
 more data and a reduction of the systematic error are needed.
%
%
\section{Summary and conclusions}
\label{sec:concl}
 The complete data samples taken by the OPAL experiment at LEP
 at $\ssee=91$~\gev and $\ssee=161-172$~\gev
 have been used to study the hadronic structure of the photon.
 \par
 A slightly better agreement between the predictions for the hadronic 
 energy flow of the various models
 and the data in the region $0.1 < \xvis < 0.6$
 is found for the data taken at $\ssee=161-172$~\gev
 than for the data collected at $\ssee=91$~\gev.
 At $\xvis<0.1$ significant differences persist, as the data 
 prefer a more pointlike hadronic energy flow 
 than assumed in either HERWIG or PYTHIA, and at 
 $\xvis>0.6$ the statistical precision is insufficient 
 to draw firm conclusions.
 \par
 Using the OPAL data at $\ssee=161-172$~\gev
 the photon structure function \ftxq has been unfolded
 as a function of $x$ in four \qsq intervals, with mean momentum transfers
 \qzm = 9, 14.5, 30 and 59~\gevsq.
 Combining the OPAL data at $\ssee = 91$~\gev and $161-172$~\gev, 
 the evolution of
 \ft with \qsq in the range $0.10 < x < 0.60$ has been measured to be
 $\ftqn=(0.16 \pm 0.05^{+0.17}_{-0.16}) +
        (0.10 \pm 0.02^{+0.05}_{-0.02})\,\ln\,(\qsq/\gevsq)$, where the first
 error is statistical and the second systematic. 
 From the OPAL data alone the slope \slop is found to be significantly 
 different from zero and consistent with the logarithmic evolution of 
 \ft with \qsq expected from QCD.
 The dependence of the scaling violation of \ft on $x$ was 
 studied in three ranges in $x$, $0.02-0.10$, $0.10-0.25$ and $0.25-0.60$.
 The QCD prediction was found to be in agreement with the data, 
 but the accuracy of the data is insufficient to show a significantly different
 slope of \ft for the three ranges in $x$ studied.
 The precision of the measurement of the logarithmic slope 
 of \ft is primarily limited by systematic errors due to the modelling 
 of the hadronic final state, not by the experimental statistics. 
 There would be scope for considerable improvements 
 if the Monte Carlo models could be improved to represent all aspects
 of the data.
 \par
 The data, over the $x$ and \qsq range studied, are equally well described
 by several of the available parton density parametrisations,
 including the GRV and SaS1D parametrisations used in this analysis.
 They are also satisfactorily described by an \ft based on the asymptotic 
 solution in leading order for three light flavours augmented by the 
 Bethe-Heitler contribution of charm quarks and a leading order hadronic 
 part based on VDM.
%
%
\par
\section*{Acknowledgements}
 We thank E.~Laenen for providing the higher order calculations for the 
 evolution of the structure function and E.~Laenen and A.~Vogt 
 for valuable discussions.
 \par
 We particularly wish to thank the SL Division for the efficient operation
 of the LEP accelerator at all energies and for
 their continuing close cooperation with
 our experimental group.  We thank our colleagues from CEA, DAPNIA/SPP,
 CE-Saclay for their efforts over the years on the time-of-flight and trigger
 systems which we continue to use.  In addition to the support staff at our own
 institutions we are pleased to acknowledge the  \\
 Department of Energy, USA, \\
 National Science Foundation, USA, \\
 Particle Physics and Astronomy Research Council, UK, \\
 Natural Sciences and Engineering Research Council, Canada, \\
 Israel Science Foundation, administered by the Israel
 Academy of Science and Humanities, \\
 Minerva Gesellschaft, \\
 Benoziyo Center for High Energy Physics,\\
 Japanese Ministry of Education, Science and Culture (the
 Monbusho) and a grant under the Monbusho International
 Science Research Program,\\
 German Israeli Bi-national Science Foundation (GIF), \\
 Bundesministerium f\"ur Bildung, Wissenschaft,
 Forschung und Technologie, Germany, \\
 National Research Council of Canada, \\
 Hungarian Foundation for Scientific Research, OTKA T-016660, 
 T023793 and OTKA F-023259.\\
%
%

%
%
\pagebreak\bigskip
%
%
%
%
%
%
%
%
%
%
\end{document}